\documentclass[%
 preprint,
 amsmath,amssymb,
 aps, physrev,
]{revtex4-2}

\usepackage{graphicx}
\usepackage{dcolumn}
\usepackage{subcaption}
\usepackage{bm}

\begin{document}

\preprint{}

\title{\textbf{Reexamination of collisional ionization cross sections including double photoionization processes} 
}%

\author{
L. Ansia$^{1,2,*}$,
P. Velarde$^{2}$,
M. Fajardo$^{1}$,
and G. O. Williams$^{1}$
}

\affiliation{
$^{1}$GoLP/Instituto de Plasmas e Fusão Nuclear, Universidade de Lisboa, 1049-001, Portugal\\
$^{2}$Instituto de Fusión Nuclear, Universidad Politécnica de Madrid, José Gutiérrez Abascal 2, 28006, Spain\\
$^{*}$Corresponding author: lucas.ansia.fernandez@tecnico.ulisboa.pt
}

\begin{abstract}
Collisional ionization (CI) cross sections in dense plasmas remain difficult to constrain due to uncertainties in plasma conditions and the overlapping spectral signatures of competing atomic processes. The use of x-ray free electron lasers (XFELs) to both heat and probe solid-density targets has significantly advanced the field by eliminating assumptions about ion density. However, questions remain regarding collisional cross sections, suprathermal electron evolution and competing atomic processes. 

In this work, we revisit experimental data from XFEL-heated aluminum, previously analyzed using collisional radiative models that did not treat the degenerate electron distribution and atomic processes self consistently. We present a new analysis using BibBarT which dynamically evolves non-thermal electron populations and explicitly includes degeneracy effects. Furthermore, we incorporate an important atomic process recently observed in plasma state that mimic signatures of CI, shake-off. Our results show that including shake-off processes improves agreement with observed emission features, and lowering recombination rates further improves the agreement with data-- indicating a possible overestimate of three-body recombination in these conditions.

\end{abstract}

\maketitle


\section{\label{sec1}Introduction}

Understanding the properties and evolution of high-energy-density (HED) plasmas is central to a wide range of fields, including inertial confinement fusion (ICF) \cite{FUS} , laboratory astrophysics \cite{ROADMAP}, and warm dense matter research \cite{WDM}. These plasmas are characterized by solid or near-solid densities and temperatures of several electronvolts, leading to highly transient, non-equilibrium conditions that are difficult to probe. While advances in experimental platforms, such as x-ray free-electron lasers (XFELs) \cite{XFEL1}, have made it possible to create and observe these extreme states with unprecedented uniformity, determining key plasma parameters—such as temperature, charge state distribution, and electron density—remains a significant challenge. 

While few studies have been able to measure the temporal evolution of XFEL produced plasmas using high-harmonic probing \cite{HHG1, HHG2}, integrated spectroscopy—and in particular the analysis of K-shell emission lines— remains one of the most effective tools for diagnosing these plasmas \cite{XFEL2}. The shapes and intensities of spectral features—such as line positions, broadening, and the presence of satellite lines—encode detailed information about the plasma's thermodynamic and charge state distribution. However, interpreting these spectra requires reliable atomic modeling under non-ideal conditions. 

A persistent source of uncertainty in spectroscopic modeling lies in the accuracy of atomic cross sections, particularly for collisional process near the continuum threshold. These rates are difficult to determine experimentally in dense plasmas, and theoretical calculations often rely on isolated-atom approximations corrected with available models of ionization potential depression (IPD) \cite{CI1, CI2}. However, IPD remains poorly constrained under strong coupled degenerate conditions, introducing additional uncertainty. As a result, most models use gas-phase atomic data adjusted by semi-empirical IPD schemes, which may not accurately capture such complex, non-ideal environments leading to incorrect collisional rates and inaccurate modeling of the spectra.

Another potential source of uncertainty in modeling dense, XFEL-driven plasmas is the treatment of the electron distribution function. Many collisional-radiative (CR) models assume a thermalized, Maxwell–Boltzmann electron population characterized by a single temperature. However, under conditions of intense photo ionization, such as those created by XFELs, the electron distribution can remain far from thermal equilibrium over timescales relevant for spectral formation \cite{NT1}. Moreover, at the solid densities and moderate temperatures typical of these experiments, Fermi–Dirac statistics must be used to describe the electron population \cite{DEGEN}. While the original analysis have attempted to incorporate these effects through non self-consistent approach \cite{OG}, the present work employs the \textsc{BigBarT} model, which evolves the non-thermal electron distribution in time while accounting for degeneracy. Details on the models and implementation can be found in refs \cite{BIGBART1, BIGBART2, BIGBART3}.

Recent studies have highlighted the importance of shake-off  (SO) in solid-density photoionized (PI) plasmas \cite{SO}. SO occurs when the sudden removal of a core electron leads to the ejection of a second electron. Unlike collisional ionization, SO is not mediated by the plasma environment and does not depend on electron temperature; rather, it is an intrinsic atomic process, with a probability determined by the initial and final electronic wavefunctions. Following a primary PI event, the SO process can create additional L-shell vacancies, giving rise to satellite emission lines indistinguishable from those produced by collisional processes. As a result, SO can mimic the signatures of hotter or more ionized conditions, leading to potential overestimation of plasma temperature or ionization degree if not properly accounted for. We found that when including SO, the spectra remain near unchanged over the entire temperature range explored (see apendix I), from the lowest up to the peak value of 13 eV, indicating a higher degree of ionization than would be expected at such relatively low intensities. Under these conditions, the spectra provide a test of the available cross sections. The creation of K-shell vacancies (via photoionization) and L-shell vacancies (via shake-off) leads to the formation of satellite lines. The persistence of these satellites is governed by the rate at which the vacancies are refilled, which, at the achieved temperatures, is dominated by three-body recombination (3BR).

In this work, we re-examine previously reported K-shell emission spectra from XFEL-irradiated aluminum \cite{OG}, with the aim of evaluating the role of shake-off processes and improving the consistency of the atomic modeling. We begin by briefly reviewing the experimental setup and conditions. The following section presents the theoretical treatment of shake-off and the methodology used to calculate the associated rates. We then revisit the assumptions made in earlier analyses—such as electron thermalization timescales and degeneracy effects—by performing self-consistent simulations. Finally, we present the full simulation results, including shake-off, and compare them with experimental data to assess the impact of each modeling approximation, including the collisional cross sections. The results point towards a re-examination of the 3BR cross sections in these conditions.

\section{\label{sec_experiment}Experiment}

The experiment was conducted at the Linac Coherent Light Source (LCLS) XFEL using the Matter in Extreme Conditions instrument \cite{LCLS1, LCLS2}. XFEL pulses of 3 keV photon energy, 35 fs duration (FWHM), and 37 $\mu$J energy were focused to a ~25 $\mu$m spot using beryllium lenses, and used to photoionize solid-density aluminum foils of 300 and 600 nm thickness.

Photo absorption of the incoming XFEL photons primarily excites K-shell electrons, producing K-holes that are predominantly filled via L-to-K Auger transitions. In aluminum, this results in the emission of ~1.4 keV Auger electrons and leaves behind two L-shell vacancies. Approximately 96$\%$  of recombination events proceed via Auger decay, while the remaining ~4$\%$ result in the emission of fluorescence photons in the soft x-ray regime, which form the measured emission spectrum.

Fluorescence was recorded using a curved crystal spectrometer with ~1 eV resolution. Detailed diagnostics of the XFEL intensity distribution, ablation pattern (F-scan), and inferred plasma conditions are described in the original work \cite{OG}. These measurements confirmed peak electron temperatures of ~13 eV in the central region of the irradiated spot.

\section{\label{sec2}Electron relaxation and degeneracy effects}

XFEL pumping generates a stream of non-thermal electrons mainly thorough photo ionization (PI) and Auger (AU) effects. The exact relaxation timescales of this process remains an open question due to its difficult experimental measure. In the original investigations, a ``time delay'' was introduced, defined as the interval between the XFEL pulse and the onset of bulk electron heating. Delays of 20--40~femtoseconds improved agreement with the data but still require a modified cross-section shape to achieve a good match with the measurements.
 
We have performed self-consistent simulations using the non-thermal model \textsc{BigBarT} to test these assumptions. Our results indicate that the actual time delay is substantially shorter, and electron bulk heating starts in few-fs time-scale as the pulse reaches the target. Recent studies \cite{BIGBART3, NT2} suggest that rapid thermalization is driven by inelastic electron–ion collisions, particularly when ionizing outer shells, closer to the continuum.

To quantify the degree of electron non-thermality, we compare the instantaneous electron entropy with that of the corresponding thermal reference distribution. We define entropy as,

\begin{equation}
    \begin{aligned}
        S = & - k_{\beta} \int g(E) \tilde{f}(E) \ln \tilde{f}(E) dE, 
    \end{aligned}
\end{equation}

where \( g(\varepsilon) =  \frac{1}{2 \pi^2} (\frac{2m}{\hbar^2})^{3/2} \sqrt{E} \) represents the density of states, with $k_{\beta}$ the Boltzman constant,  $m$ the electron mass, $\hbar$ the reduced Planck constant and \( \tilde{f}(\varepsilon) \) is the occupation factor, which varies between 0 (no electrons occupying any state) and 1 (full occupation), such that the total density is $n_e = \int f(\varepsilon) d\varepsilon = \int g(\varepsilon) \tilde{f}(\varepsilon) \, d\varepsilon$.  The corresponding entropy deviation from equilibrium, 

\begin{equation}
\Delta S = S_{\mathrm{th}} - S
\end{equation}
where $S$ is the entropy of the instantaneous electron energy distribution and $S_{\mathrm{th}}$ is the entropy of a thermal distribution with the same density and mean energy. This dimensionless quantity vanishes for a purely thermal distribution and increases as the electron population departs from thermal equilibrium.

\begin{figure}
    \centering
    \includegraphics[scale=1]{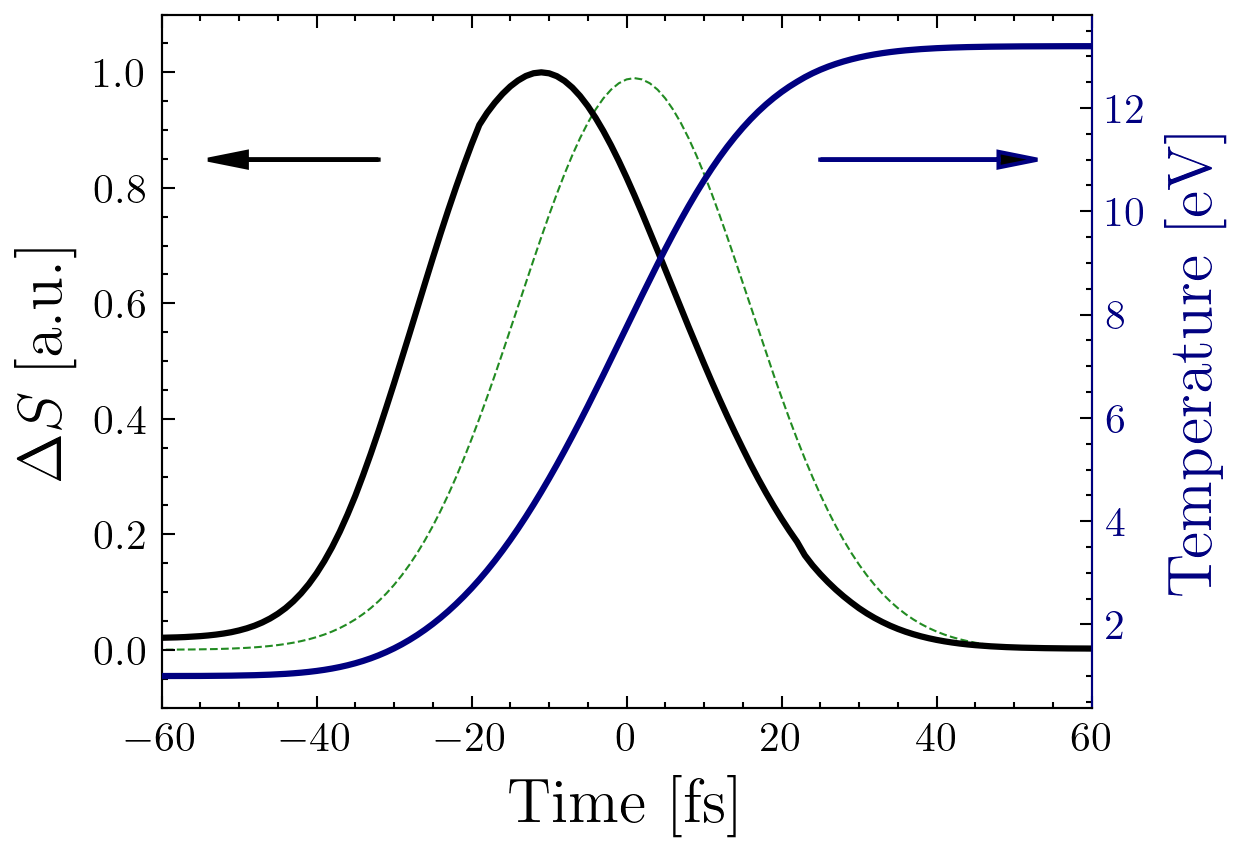}
    \caption{\label{entropy} Temporal evolution of the non-thermal entropy $\Delta S$ (solid black line) and the bulk electron temperature (solid blue line). The dashed curve shows the XFEL pulse temporal profile (Gaussian envelope), centered at $t=0$.
}
\end{figure}

Figure \ref{entropy} shows the temporal evolution of $\Delta S$ (black solid line), together with the XFEL pulse envelope (green dashed line) and the bulk electron temperature (blue line). An increase of $\Delta S$ is observed as the irradiation begins, followed by a rapid decay on a few-femtosecond timescale, showing that the energy deposited by the pulse is quickly transferred to the bulk electron population. This peak corresponds to the formation of a pronounced non-thermal feature driven 3BR, which progressively disappears as the electron distribution heats up and elastic collisions smooth it out \cite{BIGBART3}.

We also include the effects of electronic degeneracy in our simulations~\cite{BIGBART3}, which further influence the ionization balance. The resulting changes stem from two main factors. First, degeneracy alters the standard Saha equilibrium. In the classical high-temperature limit, the electron chemical potential \(\mu\) satisfies

\begin{equation}
e^{-\mu/kT} = \frac{n_e}{2}\left(\frac{h^2}{2\pi m kT}\right)^{3/2},
\end{equation}

\noindent
where T is the electron temperature and K is the Boltzmann constant. Under degenerate conditions, this relation breaks down. Instead, \(\mu\) must be determined self-consistently by solving Fermi-Dirac integrals to conserve the total electron density. 

\begin{equation}
    n = \frac{(2m_e)^{3/2} T^{3/2}}{2\pi^2 \hbar^3} F_{1/2}(\eta) = \frac{(2m_e)^{3/2} T^{3/2}}{2\pi^2 \hbar^3} \int_0^\infty \frac{x^j}{e^{x - \eta} + 1} dx.
\end{equation}

\noindent
where \( \eta = \mu / T \) is the reduced chemical potential.

\begin{figure}
    \centering
    \includegraphics[scale=1]{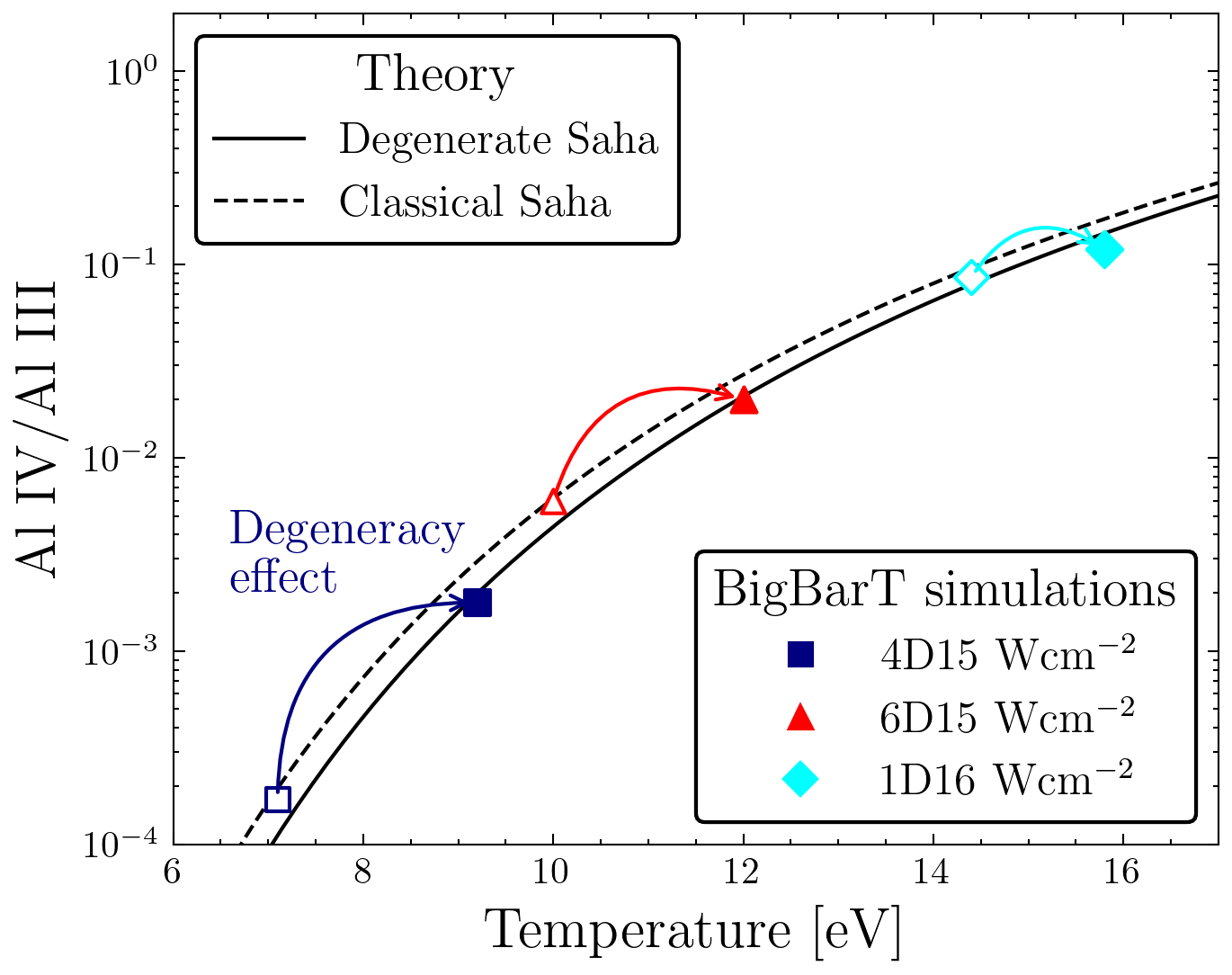}
    \caption{\label{fsaha} The ratio of ion populations, Al IV/ Al III, plotted relative to the thermal electron temperature, plotted for classical and degenerate treatments of the Saha equation. Open and filled markers show \textsc{BigBarT} simulation results for classical and degenerate cases, respectively, at three different intensities.
}
\end{figure}

Under strong coupling conditions, the chemical potential shifts to higher (positive) values, indicating that newly added particles occupy states at or near the Fermi level, given the exclusion principle. This reduced state availability restricts collisional pathways, leading to slightly lower ionization under degenerate conditions, as shown in Fig.~\ref{fsaha}.

Second, a degenerate free-electron gas follows a Fermi--Dirac distribution, which reduces its specific heat compared to a classical Maxwell--Boltzmann gas, as can be seen from the Sommerfeld expansion. This is because most electrons are Pauli-blocked below the Fermi energy, and only a small fraction near the Fermi surface can absorb energy and be thermally excited. Consequently, the number of active degrees of freedom is strongly reduced, leading to a lower heat capacity. In thermodynamic terms, $C_V = T (\partial S / \partial T)$, so the limited number of accessible electronic states constrains the entropy increase with temperature. As a result, for a given absorbed energy, a degenerate electron gas exhibits a larger temperature increase due to its reduced heat capacity. This is the regime considered here, as the XFEL photon energy is well above the K-edge, so Pauli blocking does not significantly affect the absorption and both systems take up nearly the same energy.

The combination of these two effects—reduced equilibrium ionization but enhanced heating—leads overall to a net increase in ionization under degenerate conditions. However, even with degeneracy included, the ionization levels remain below those required to match the experimental spectra.

These effects are illustrated in Fig.~\ref{fsaha}, which shows theoretical Saha equilibrium curves alongside simulation results for three different XFEL intensities.

\section{\label{sec3}Shake off process}
\subsection{Theory}

Shake-off is an atomic process that occurs following a primary electron ionization, in which the sudden removal of a core electron leads to the ejection of an additional, more weakly bound electron. This process arises from the abrupt change in the potential experienced by the remaining electrons and is therefore governed purely by atomic structure, independent of the surrounding plasma conditions. As a result, SO can contribute a consistent background of additional ionization across a wide range of densities and temperatures, particularly in photoionized plasmas. It may produce spectral features similar to those arising from collisional ionization, which can lead to misinterpretation of plasma parameters if not properly accounted for. In order to evaluate the contribution of SO to the observed emission, we calculate the probability of this process occurring as a follow-up to photoionization.

In order to calculate the probability of a shake process occurring, sudden perturbation theory (SPT) is commonly employed. This theory is based on the idea that a quantum system is initially in a state \( | \Psi \rangle \), corresponding to a Hamiltonian \( \mathcal{H^-} \), which undergoes a sudden transition to a new Hamiltonian \( \mathcal{H^+} \). This transition happens so rapidly that the system does not have time to fully adjust or rearrange itself. The initial state of the system can then be expanded in terms of the new basis as:

\begin{equation}
  | \Psi \rangle = \sum_{i, k} \langle \psi^+_k | \psi^-_i \rangle | \psi^+_k \rangle  
\end{equation}

where \( | \psi^-_i \rangle \) and \( | \psi^+_k \rangle \) denote the eigenstates of the old Hamiltonian \( \mathcal{H} \) and the new Hamiltonian \( \mathcal{H^+} \), respectively. From this, the overlap \( \left| \langle \psi^+_k | \psi^-_i \rangle \right|^2 \) represents the probability that the state \( | \psi^-_i \rangle \) will transition to the new state \( | \psi^+_k \rangle \).

The probability for an electron to remain in the initial state \( nlj \) is:

\begin{equation}
   S = \left| \langle \psi^+_{nlj} | \psi^-_{nlj} \rangle \right|^{2} 
   = \left| \int \psi^{+ *} _{nlj}\, \psi^-_{nlj} \, dr \right|^{2}
\end{equation}

For a shell of \( N \) electrons, the probability of the entire shell remaining in the initial level is \( P_{O} = S^N \). For the shake-up process, where monopole transitions are involved, selection rules only allow changes in the principal quantum number. For a shell of \( N \) electrons, the probability of a transition from the state \( nlj \) to \( n'lj \) is:

\begin{equation}
   P_{SU}  = N \left| \langle \psi^+_{n'lj} | \psi^-_{nlj} \rangle \right|^{2} S^{N-1} = N \left| \int \psi^{+ *} _{n'lj} \psi^-_{nlj} \, dr \right|^{2} S^{N-1}
\end{equation}

For the shake-off process, the procedure is similar, but we must account for all possible continuum wave functions \( | \psi_{\epsilon lj} \rangle \):

\begin{equation}
   P_{SO} = N \left| 
   \int d\epsilon \int \psi^{+ *}_{\epsilon lj} \, \psi^{-}_{nlj}\, dr 
   \right|^{2} S^{N-1}
\end{equation}

This calculation can be computationally expensive. To address this, we adopt the method developed by Carlson and Nestor \cite{SO1}, which exploits the fact that the final state either remains in the same configuration or undergoes one of two shake processes. In addition, transitions to already occupied states are forbidden by the Pauli exclusion principle; their associated probability is denoted by \( P_{F} \) and given by , 

\begin{equation}
    P_{F} = \sum_{n'} N \frac{N'}{2j+1} \left| \int {\psi^{+ *}_{n'lj}} \psi^-_{nlj} \, dr \right|^{2} S^{N-1},
\end{equation}

where \( N' \) is the number of electrons already occupying the final subshell \( n'lj \), which has a total degeneracy of \( 2j+1 \). Thus, the factor \( \frac{N'}{2j+1} \) represents the fraction of sublevels that are already occupied and therefore forbidden by the Pauli exclusion principle. With this definition, the shake-off probability can be written as

\begin{equation}
    P_{SO} = 1 - (P_{0} + P_{SU} + P_{F}).
\end{equation}

Once \( P_{O} \), \( P_{SU} \), and \( P_{F} \) are known, the shake-off probability \( P_{SO} \) is easily obtained.

In the WDM regime, ionization potential depression (IPD) should be included when evaluating the bound states relevant for shake-up and shake-off processes. While the exact magnitude of IPD remains imperfectly constrained under strongly coupled, degenerate conditions, the present model is mainly sensitive to the distinction between states treated as bound and those regarded as pressure-ionized. In isolated atoms, shake-up is usually more probable because the overlap between bound orbitals is strongest for energetically nearby states. Under the plasma conditions considered here, however, the n=3 states of Al can be taken as pressure-ionized, consistent with the fact that aluminum contributes three valence electrons to the plasma. We therefore retain only levels up to the 2p shell as bound. Consequently, shake-up contributions are strongly suppressed, and the dominant contribution is shifted toward shake-off processes \cite{SO2}.

\section{Results}

\begin{figure}
    \centering
    \includegraphics[scale=0.5]{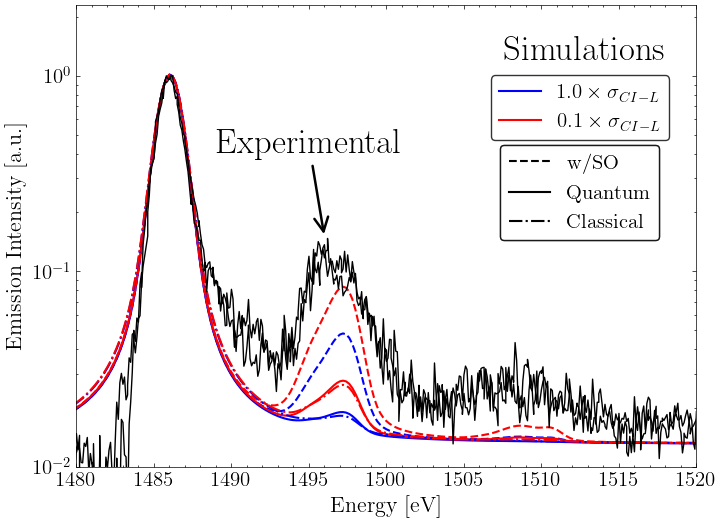}
    \caption{\label{result} 
    Emission spectra as a function of photon energy for two different collisional cross-section scalings (\(1.0\times\sigma_{CI-L}\), and \(0.1\times\sigma_{CI-L}\)).  
Results are shown for classical (Maxwell–Boltzmann), degenerate (Fermi–Dirac), and degenerate including shake-off (SO) process.  
}
\end{figure}

We now compare our self-consistent simulations, which include non-thermal electron dynamics and electronic degeneracy within the \textsc{BigBarT} model, to the experimental emission spectra. Each simulation result includes a set of six intensities to match the fluence scan described in the original article \cite{OG} (more details in Appendix \ref{sec_apendix}). Results with and without the inclusion of the shake-off (SO) process are presented in Fig. \ref{result}.

Even with all processes included, the simulated spectra underestimate the intensity of the L-shell satellite features (particularly the K1L7–K2L6 transitions) observed experimentally. This is notable because the SO rates calculated here are sufficient, in principle, to account for the observed satellite intensity ratio.

However, the contribution from shake-off is strongly suppressed by the rapid refilling of L-shell vacancies through electron recombination. Under the present plasma conditions, 3BR dominates this process. This suggests a possible overestimation of the collisional cross sections used to model these rates. Experimental measurements of these cross sections remain extremely challenging, and first-principles calculations—particularly those not relying on ad hoc ionization potential depression (IPD) models—are still subject to considerable uncertainty. Nagler et al., made solid aluminum transparent through saturable absorption using 15 fs X-ray pulses at LCLS \cite{SATURABLE}, suggesting that core-hole lifetimes are on the order of tens of femtoseconds. The standard cross-sections employed here \cite{CI2} correspond to lifetimes of a few femtoseconds, suggesting a divergence in the observed and calculated rates of 3BR in these conditions.

To explore the sensitivity of the emission spectra to these rates, we performed simulations in which the collisional cross sections were reduced by an order of magnitude. This adjustment simultaneously lowers both collisional ionization and 3BR rates, consistent with Fowler's micro-reversibility relations. The resulting spectra (red lines in Fig. \ref{result} ) exhibit significantly improved agreement with experimental observations, primarily due to the reduced refilling of L-shell holes, which allows shake-off pathways to contribute more effectively.

Interestingly, even in the absence of shake-off, lowering the collisional cross sections leads to a closer match with the experimental satellite features. While this may seem counterintuitive, the effect is temporal: reducing the 3BR rate delays L-shell hole refilling, allowing configurations such as K2L7 to persist longer and peak closer to the XFEL pulse maximum. Essentially as L-shell CI is low in this regime of temperature, decreasing the CI cross section only affects the emission spectrum by decreasing it's counter process, three-body recombination.

While it remains difficult to place tight constraints on the absolute values of collisional cross sections under these plasma conditions, our results suggest that cross sections based on IPD-scaled atomic models may overestimate 3BR recombination rates. Further experimental and theoretical work is needed to benchmark these quantities in solid-density, XFEL-driven plasmas.

\section{\label{sec_conclusions}Conclusions}
We have shown that shake-off processes can play an important role in shaping the $K\alpha$ spectra emitted by plasmas under extreme conditions. By incorporating shake and degeneracy effects within a single self-consistent atomic model, we revisit experimental data obtained prior to the availability of such models. This approach allows us to isolate and quantify separately the effects of shake processes, degeneracy, and collisional ionization.

We find that shake-off processes contribute significantly to the satellite peaks, which are important for plasma diagnostics. Nevertheless, the total satellite emission depends strongly on the 3B recombination rate, making it sensitive to the choice of cross sections. We find that the standard values reported in the literature lead to an underestimation of the hole lifetime; only by decreasing 3B cross section we do obtain agreement with the experimental data.

While it remains difficult to place tight constraints on the absolute values of collisional cross sections under these plasma conditions, our results suggest that cross sections based on IPD-scaled atomic models may overestimate 3BR recombination rates. At the same time, good agreement with standard collisional cross sections has been reported in a resonant XFEL-pumping experiment on magnesium, which reached electron temperatures of about 200 eV. Further experimental and theoretical work is therefore needed to benchmark these quantities in solid-density, XFEL-driven plasmas.

\begin{acknowledgments}

L. A. acknowledge support from by FCT (Foundation for Science and Technology - Portugal) under Grant No. UI/BD/153734/2022.

This project has received funding from the European Union's Horizon Europe programme, European Research Council EIC Pathfinder Open under grant agreement No 101047223 (NanoXcan).

The authors gratefully acknowledge the Universidad Politécnica de Madrid (www.upm.es) for providing computing resources on Magerit Supercomputer.

\end{acknowledgments}

\bibliography{apssamp}

\appendix
\section{\label{sec_apendix}Intensity dependence}

Within the 0D modeling framework, a more realistic representation of the total emitted spectrum is obtained by considering several intensity zones and averaging their individual contributions. This provides an approximate description of the spatial XFEL intensity distribution across the focal spot, which cannot be captured in a single-zone calculation. In this Appendix, we present the six intensity values included in this procedure, together with the corresponding spectra and electron-temperature evolution. A key result is that, in the presence of shake-off (SO), the spectra vary only weakly across this intensity range, since the satellite feature is already present and remains largely insensitive to the plasma temperature. Each panel is associated with one of the two collisional cross-section scalings considered in the main text, $1.0\times\sigma_{CI-L}$ and $0.1\times\sigma_{CI-L}$, and corresponds to one of the spectra presented in the Results section. Results are shown for the classical (Maxwell--Boltzmann), degenerate (Fermi--Dirac), and degenerate cases including SO.

\begin{figure*}[t]
    \centering

    \begin{subfigure}{0.48\textwidth}
        \centering
        \includegraphics[width=\linewidth]{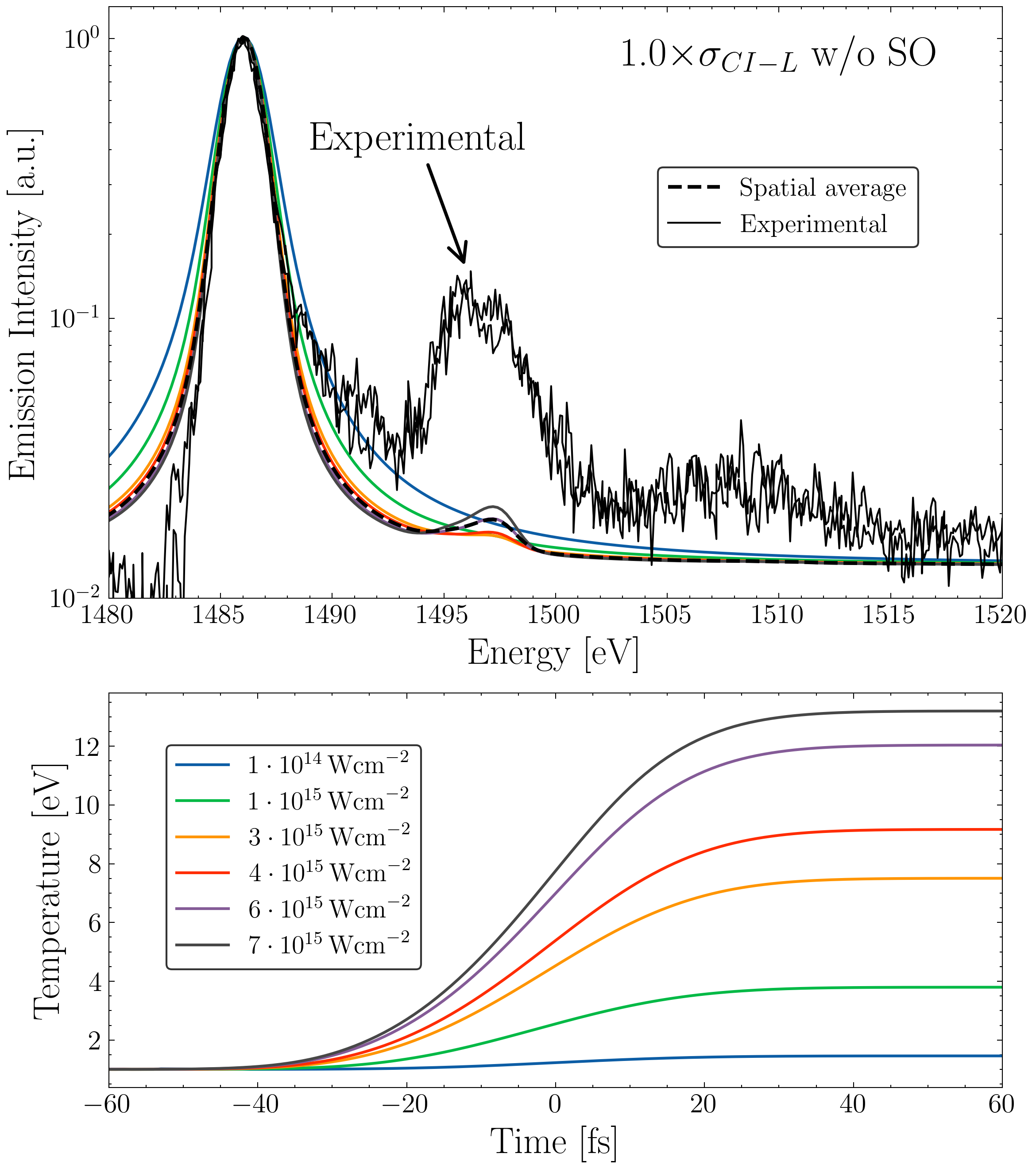}
        \caption{Spectra and electron-temperature evolution obtained for the six intensity values included. This set corresponds to the case with collisional cross-section scaling $1.0\times\sigma_{CI-L}$, no shake-off (SO) and FD statistics}
        \label{fig:1a}
    \end{subfigure}
    \hfill
    \begin{subfigure}{0.48\textwidth}
        \centering
        \includegraphics[width=\linewidth]{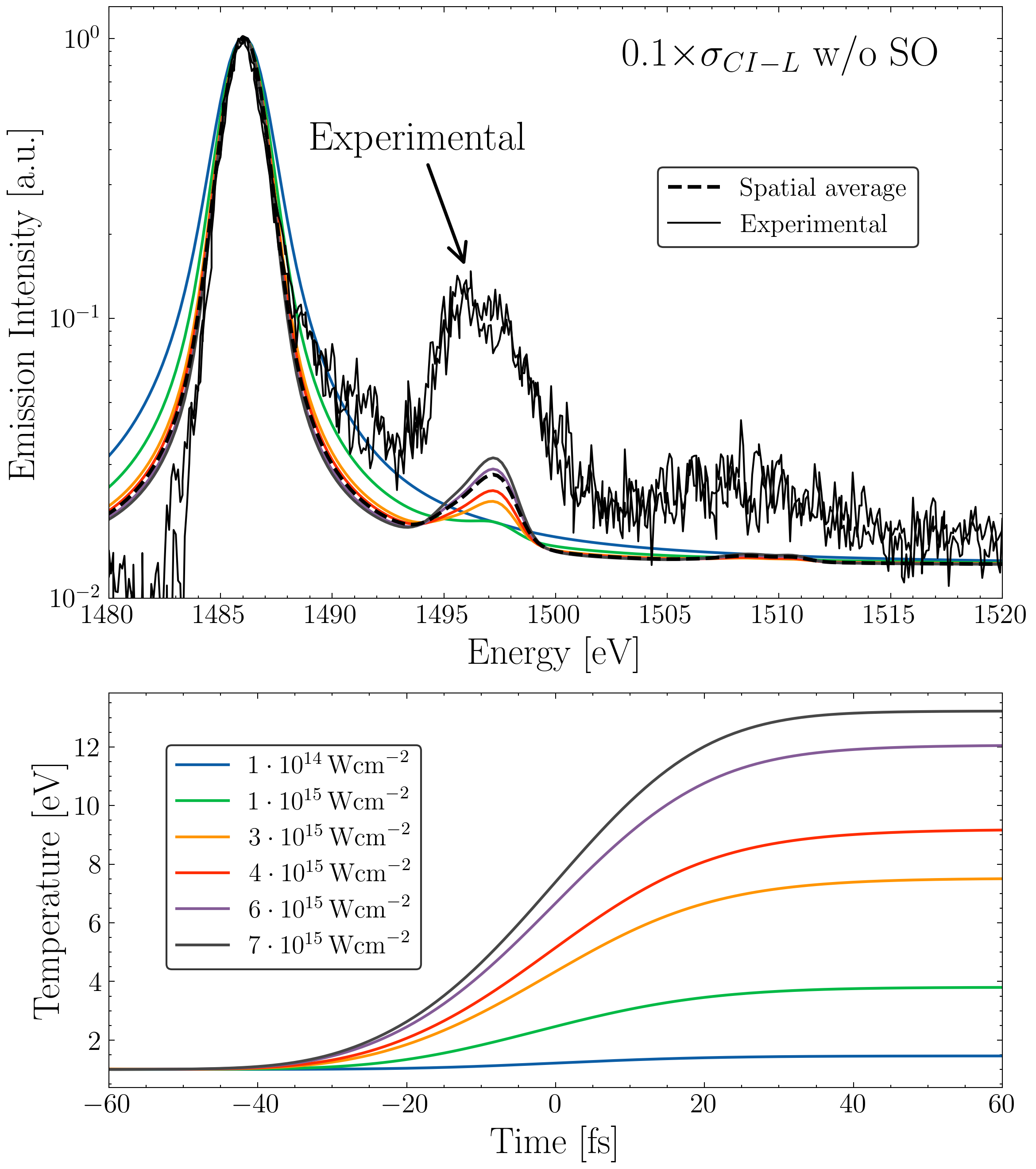}
        \caption{Spectra and electron-temperature evolution obtained for the six intensity values included. This set corresponds to the case with collisional cross-section scaling $0.1\times\sigma_{CI-L}$, no shake-off (SO) and FD statistics}
        \label{fig:1b}
    \end{subfigure}

\end{figure*}

\begin{figure*}[t]
    \centering

    \begin{subfigure}{0.48\textwidth}
        \centering
        \includegraphics[width=\linewidth]{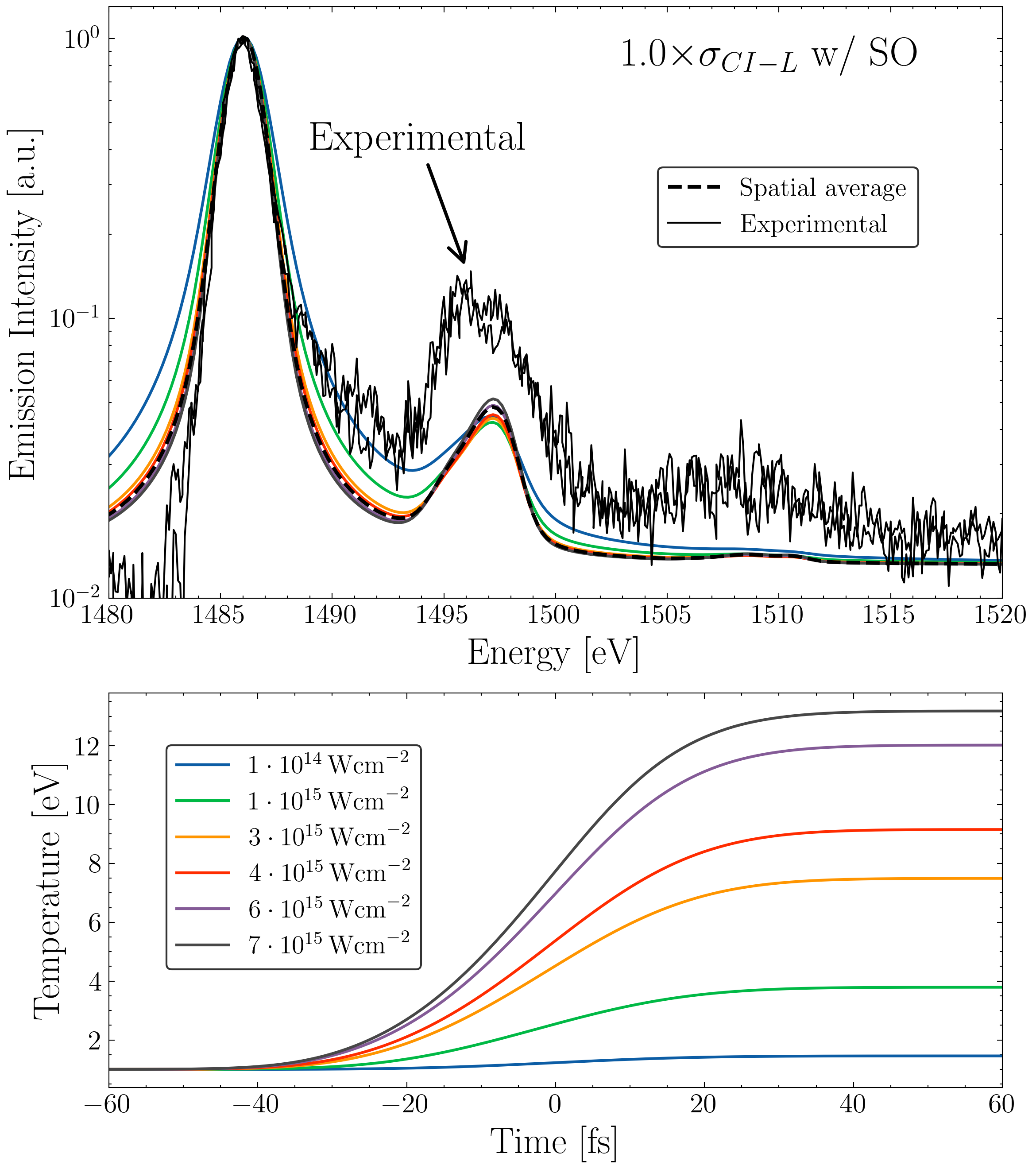}
        \caption{Spectra and electron-temperature evolution obtained for the six intensity values included. This set corresponds to the case with collisional cross-section scaling $1.0\times\sigma_{CI-L}$ including shake-off (SO) and FD statistics}
        \label{fig:1c}
    \end{subfigure}
    \hfill
    \begin{subfigure}{0.48\textwidth}
        \centering
        \includegraphics[width=\linewidth]{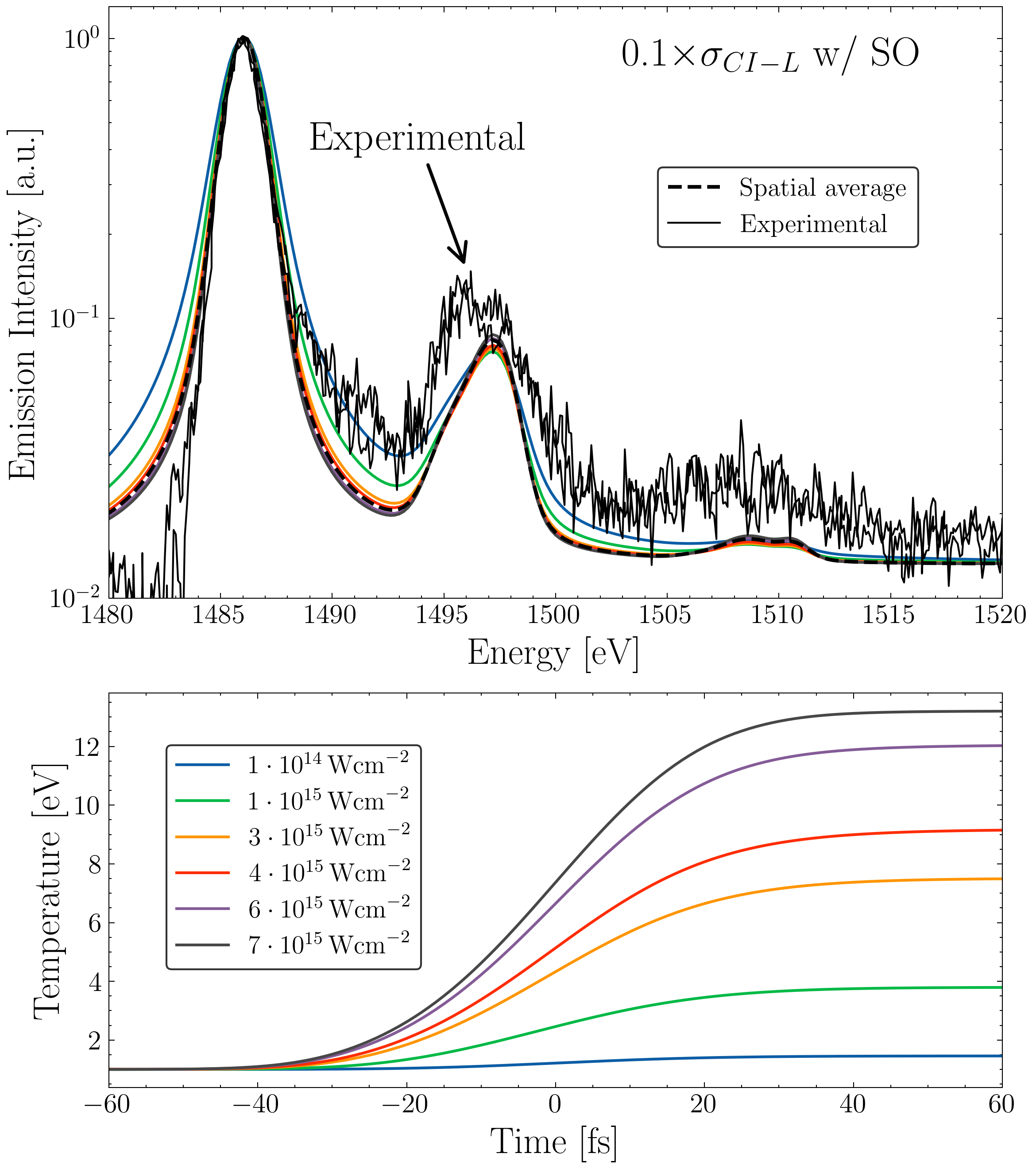}
        \caption{Spectra and electron-temperature evolution obtained for the six intensity values included. This set corresponds to the case with collisional cross-section scaling $0.1\times\sigma_{CI-L}$ and including shake-off (SO) and FD statistics}
        \label{fig:1d}
    \end{subfigure}

\end{figure*}

\begin{figure*}[t]
    \centering

    \begin{subfigure}{0.48\textwidth}
        \centering
        \includegraphics[width=\linewidth]{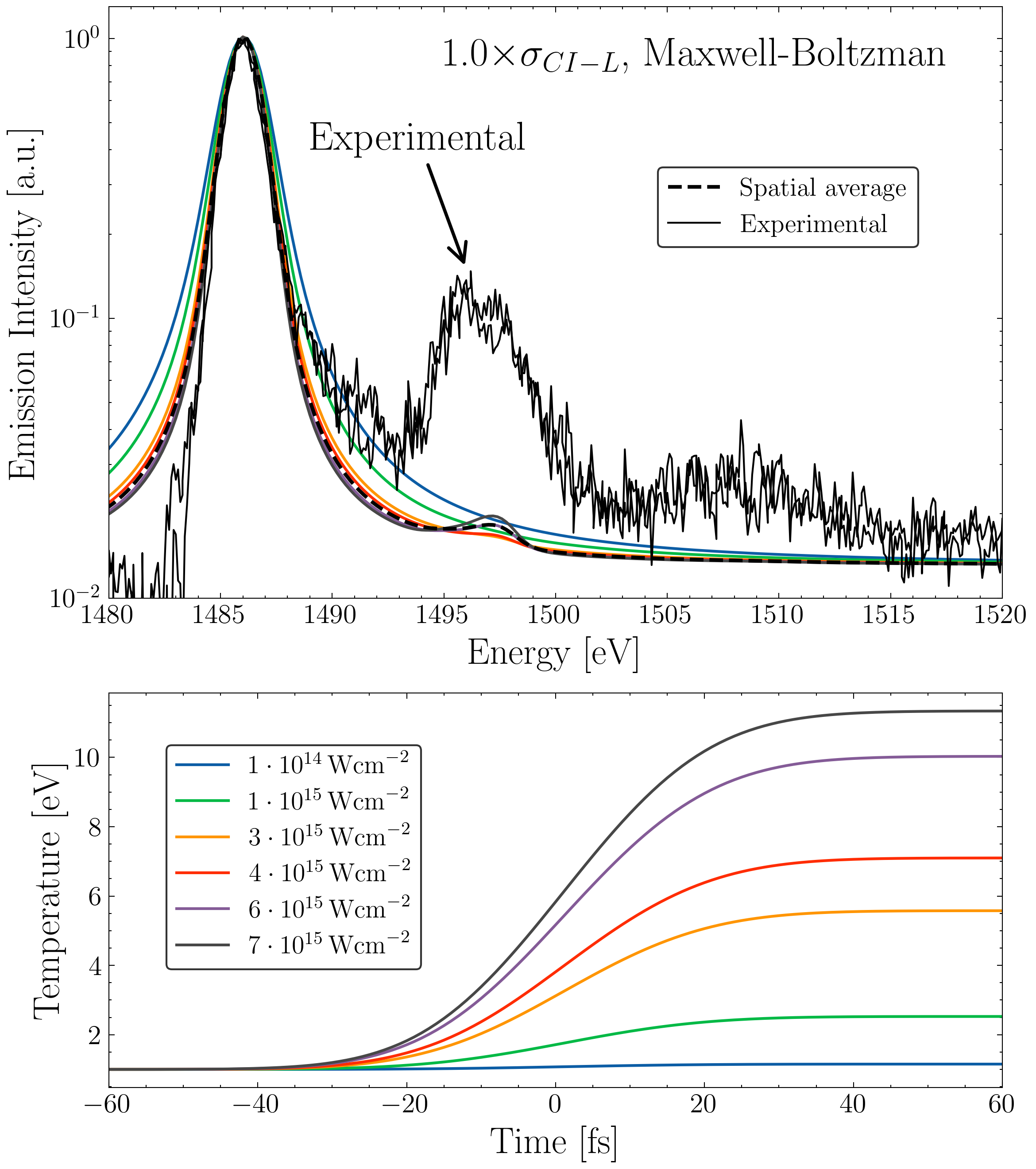}
        \caption{Spectra and electron-temperature evolution obtained for the six intensity values included. This set corresponds to the case with collisional cross-section scaling $1.0\times\sigma_{CI-L}$ including shake-off (SO) and FD statistics}
        \label{fig:1e}
    \end{subfigure}
    \hfill
    \begin{subfigure}{0.48\textwidth}
        \centering
        \includegraphics[width=\linewidth]{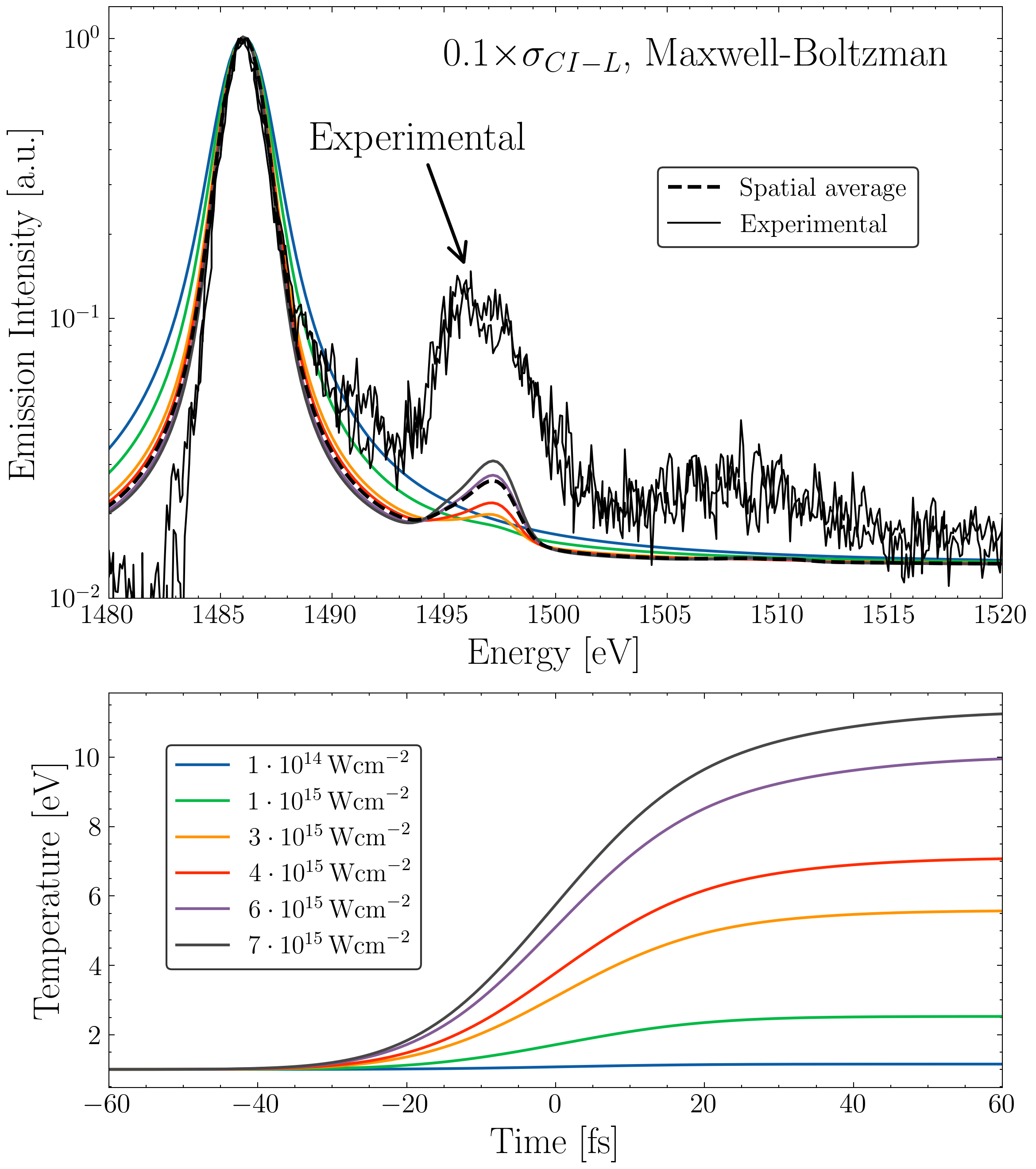}
        \caption{Spectra and electron-temperature evolution obtained for the six intensity values included. This set corresponds to the case with collisional cross-section scaling $0.1\times\sigma_{CI-L}$ without shake-off (SO) and using MB statistics}
        \label{fig:1f}
    \end{subfigure}

\end{figure*}


\end{document}